\documentclass[preprint,aps,amsmath,amssymb,superscriptaddress]{revtex4}

\usepackage{graphicx}% Include figure files
\usepackage{dcolumn}% Align table columns on decimal point
\usepackage{bm}% bold math

\begin{document}

%\preprint{APS/123-QED}

\title{Quantum simulation of Fermi-Hubbard models in semiconductor quantum dot arrays}
\author{Tim Byrnes}
\affiliation{National Institute of Informatics, 2-1-2
Hitotsubashi, Chiyoda-ku, Tokyo 101-8430, Japan}
\affiliation{Institute of Industrial Science,
University of Tokyo, 4-6-1 Komaba, Meguro-ku, Tokyo 153-8505,
Japan} 

\author{Na Young Kim}
\affiliation{E. L. Ginzton Laboratory, Stanford University,
Stanford, CA 94305}

\author{Kenichiro Kusudo}
\affiliation{National Institute of Informatics, 2-1-2
Hitotsubashi, Chiyoda-ku, Tokyo 101-8430, Japan}

\author{Yoshihisa Yamamoto}
\affiliation{National Institute of Informatics, 2-1-2
Hitotsubashi, Chiyoda-ku, Tokyo 101-8430, Japan}
 \affiliation{E.
L. Ginzton Laboratory, Stanford University, Stanford, CA 94305}
\date{\today}% It is always \today, today,

\begin{abstract}
We propose a device for studying the Fermi-Hubbard model with long-range Coulomb interactions using an array of quantum dots defined in a semiconductor two-dimensional electron gas system. Bands with energies above the lowest energy band are used to form the Hubbard model, which allows for an experimentally simpler realization of the device. We find that depending on average electron density, the system is well described by a one- or two-band Hubbard model. Our device design enables the control of the ratio of the Coulomb interaction to the kinetic energy of the electrons independently to the filling of the quantum dots, such that a large portion of the 
Hubbard phase diagram may be probed. Estimates of the Hubbard parameters suggest that a metal-Mott insulator quantum phase transition and a $d$-wave superconducting phase should be observable using current fabrication technologies. 
\end{abstract}

\maketitle

Along with quantum computers, quantum simulators promise to offer performance
exceeding what is possible using only classical physics \cite{feynman82}.  A quantum 
simulator is a purpose-built device that simulates a particular
quantum many-body problem that is intractable on a classical
computer. Several experiments and theoretical studies have indicated that such a device is possible to build 
\cite{vanderzant92,greiner02,hofstetter02,micheli06,hartmann06,angelakis07,greentree06,byrnes07}. 
Here we show that a similar device can be built
for the fermionic Hubbard model using a semiconductor quantum dot (QD) array system. 
The two-dimensional Hubbard model is particularly interesting as it is one of the central models used to describe strongly-correlated phenomena such as 
metal-insulator transitions \cite{imada98}, magnetism \cite{tasaki98}, and high-temperature superconductivity \cite{moriya03}. 
Despite decades of intensive research, a complete solution remains unavailable due to difficulties in numerical and analytic methods. 

\begin{figure}[t]
\scalebox{0.6}{\includegraphics{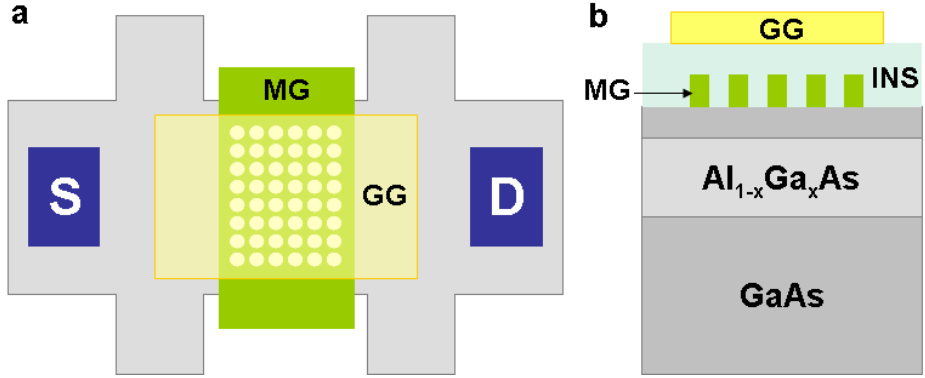}}
%\scalebox{0.45}{\includegraphics[width=\textwidth, bb=0 0 761 321]{figure_1.png}}
\caption{\label{fig:device} (color online) (a) A schematic top view of the proposed
device.  A two-dimensional Schottky mesh gate (MG) is
patterned in the central region, and a top global gate (GG) is placed
on top of the MG separated by an insulating layer such as Si$_3$N$_4$ (INS). 
Source (S) and drain (D) ohmic contacts in the Hall bar mesa
structure access the two-dimensional
electron gas (2DEG) system. (b) Cross-sectional view of the device. The 2DEG
is formed at the interface between AlGaAs and GaAs by applying a
voltage to the GG. The periodic potential is created by the 2D MG. }
\end{figure}

The type of device we consider is an undoped GaAs/AlGaAs heterostructure, 
with a two-dimensional electron gas system (2DEG) formed at the interface 
(Fig. \ref{fig:device}). 
The 2DEG is formed by applying a positive voltage to a metallic top gate 
(the ``global gate'' (GG)) \cite{willett07}. Our choice of an undoped system, as opposed to a 
modulation doped system, is important for a clean realization of the Hubbard model so that 
impurities in the system are reduced to a minimum.   In addition to the 
GG, a 2D mesh gate (MG) is patterned over a large area 
(e.g. 30 $\mu$m $\times $ 30 $\mu$m ) of the device. Applying a voltage to the MG 
induces an in-situ tunable periodic lattice potential, such that an array of coupled QDs is 
created. The MG is separated by an electrically insulating layer from the GG, 
so that the average electron number in the QDs can be controlled 
independently of the inter-QD coupling \cite{willett07}. 
The periodicity of the MG considered in this paper is $ \lambda \approx 0.1$  $\mu$m, 
which is an experimentally achievable size using current lithography techniques. The 2DEG is 
formed at a relatively shallow position relative to the surface (e.g. a depth $ d \approx 30$ nm). 
This is advantageous in order to achieve sharp QD trapping potentials, as well as reducing the 
overall Coulomb repulsion between electrons in a single QD via electrical screening from the metal 
gates. Without screening, the Coulomb interaction is typically much larger than the kinetic energy 
$ \frac{e^2}{4 \pi \epsilon \lambda} \gg \frac{\hbar^2}{2 m^* \lambda^2} $ in our semiconductor system 
($ \lambda = 0.1$  $\mu$m, $ \epsilon = 13 \epsilon_0 $ is the 
permittivity, $ m^* = 0.067 m_e $ is the effective electron mass in GaAs). The screening effect 
allows access to an interesting regime of the Hubbard model where quantum phase transition (QPT) 
phenomena are expected to occur, with the 
Coulomb repulsion $ U $ and the nearest neighbor hopping $ t $ being on the same order.

One of the key 
features of our device is that we consider average electron numbers beyond the occupation of the 
first band formed by the periodic potential. Typical high-mobility semiconductor samples
have electron densities in the region of  $\sim 10^{11}$ $\mbox{cm}^{-2}$, corresponding to 10 electrons per 
QD for a periodicity of $ \lambda = 0.1 $ $\mu$m. Past studies \cite{byrnes07,hofstetter02} have
assumed the occupation of only the lowest energy band of the imposed periodic potential, which corresponds to 
low electron densities where it is difficult to achieve high mobilities due to the presence of impurities. 
Low-density 2DEGs have also been experimentally observed to undergo a metal-insulator-like transition as
a function of electron density \cite{sarma05}. In order not to mask the effects of the effective Hubbard 
model due to such low-density effects it is advantageous to work in a high electron mobility 
regime where the system is in an unambiguous metallic state.  

Due to the presence of the electrons in the lower energy bands,
the Coulomb interaction between two electrons in the higher energy bands will experience a screening effect. 
In the standard procedure of obtaining an effective Hubbard model, one usually ignores the presence of the 
core electrons and considers only the outermost filled band of the periodic array \cite{tasaki98}. 
In order to take into account of this screening effect at a quantum mechanical level, it is necessary to incorporate correlations between electrons in the outer orbitals and the core electrons. We can take this screening effect 
into account by the following procedure. First, let us model the 2DEG electrons in a periodic potential 
$ V_M (\bm{x}) = V_0 \left[ \cos (2 \pi x / \lambda) + \cos(2 \pi y / \lambda) \right] $ by a very general
multi-Hubbard  model
\begin{eqnarray}
\label{fulllatticehamiltonian}
H & = & \sum_{\sigma \: \bm{n} \: \bm{n}' \: \bm{j} \: \bm{j}'} {\cal T} (\bm{n},\bm{n}',\bm{j},\bm{j}') c_{\bm{j} \bm{n} \sigma}^\dagger c_{\bm{j}' \bm{n}' \sigma}
\nonumber \\
&+ & \frac{1}{2} \sum_{\shortstack{\footnotesize $\sigma \: \sigma'$ \\ \footnotesize  $\bm{n}_1 \: \bm{n}_2 \: \bm{n}_3 \: \bm{n}_4 $ \\ \footnotesize $\bm{j}_1 \: \bm{j}_2 \: \bm{j}_3 \: \bm{j}_4 $}}
{\cal U} (\bm{j}_1,\bm{j}_2,\bm{j}_3,\bm{j}_4,\bm{n}_1,\bm{n}_2,\bm{n}_3,\bm{n}_4)  c_{\bm{j}_1 \bm{n}_1 \sigma}^\dagger 
c_{\bm{j}_2 \bm{n}_2 \sigma'}^\dagger c_{\bm{j}_3 \bm{n}_3 \sigma'} c_{\bm{j}_4 \bm{n}_4 \sigma},
\end{eqnarray}
where $ c_{\bm{j} \bm{n} \sigma} $ is the fermion annihilation operator
associated with the site $ \bm{j} = (j_x , j_y)  $, band $ \bm{n}  =(n_x , n_y) $ (where $n_x,n_y\ge 1$), and spin $ \sigma $.  For each band $ \bm{n} $ we may define
a Wannier basis $ w_{\bm{n}} (\bm{x}) $, from which we may define the hopping $ {\cal T} (\bm{n},\bm{n}',\bm{j},\bm{j}') = \int d^2 x w_{\bm{n}}^* (\bm{x}-\bm{x}_{\bm{j}}) H_0 (\bm{x}) w_{\bm{n}'} (\bm{x}-\bm{x}_{\bm{j}'})  $ and Coulomb $ {\cal U} (\bm{j}_1,\bm{j}_2,\bm{j}_3,\bm{j}_4,\bm{n}_1,\bm{n}_2,\bm{n}_3,\bm{n}_4) = \int d^2 x \int d^2 x'  w_{\bm{n}_1}^* (\bm{x}'-\bm{x}_{\bm{j}_1}) w_{\bm{n}_2}^* (\bm{x}-\bm{x}_{\bm{j}_2}) 
U_C (\bm{x},\bm{x}') w_{\bm{n}_3} (\bm{x}-\bm{x}_{\bm{j}_3}) w_{\bm{n}_4} (\bm{x}'-\bm{x}_{\bm{j}_4}) $ matrix elements,
where the 
single electron Hamiltonian is $ H_0 ( \bm{x} ) = - \frac{\hbar^2}{2m^*} \nabla^2 +  V_M (\bm{x}) $. 
Due to the presence of the metal gates at the surface, we use a screened Coulomb interaction 
$ U_C (\bm{x},\bm{x}')  = e^2 f_{\mbox{\tiny s}} (\bm{x},\bm{x}') / 4 \pi \epsilon |\bm{x}-\bm{x}'|$, where 
$ f_{\mbox{\tiny s}} (\bm{x},\bm{x}') = 1- |\bm{x}-\bm{x}'|/\sqrt{(x-x')^2 +
(y-y')^2+(z+z'+2d)^2} $.  The form of the screening function $ f_{\mbox{\tiny s}} (\bm{x},\bm{x}') $
is derived using the method of images by assuming a uniform metal plate at the surface. 

Let us now split the Hamiltonian (\ref{fulllatticehamiltonian}) into two parts: one corresponding to ``on-site'' 
terms
\begin{eqnarray}
 H_{\mbox{\tiny on-site}} & = & \sum_{\sigma \: \bm{n} \: \bm{j} } \epsilon_{\bm{n}} c_{\bm{j} \bm{n} \sigma}^\dagger c_{\bm{j} \bm{n} \sigma} + \frac{1}{2} \sum_{\shortstack{\footnotesize $\sigma \: \sigma' \: \bm{j} $ \\ \footnotesize  $\bm{n}_1 \: \bm{n}_2 \: \bm{n}_3 \: \bm{n}_4 $ }} {\cal \tilde{U}}_{\bm{n}_1,\bm{n}_2,\bm{n}_3,\bm{n}_4} c_{\bm{j} \bm{n}_1 \sigma}^\dagger 
c_{\bm{j} \bm{n}_2 \sigma'}^\dagger c_{\bm{j} \bm{n}_3 \sigma'} c_{\bm{j} \bm{n}_4 \sigma}
\label{onsitehamiltonian}
\end{eqnarray}
and the other corresponding to all the remaining terms $ H_{\mbox{\tiny site-site}} \equiv H - H_{\mbox{\tiny on-site}} $. All terms in
$ H_{\mbox{\tiny site-site}} $ contain operators connecting two sites $ \bm{j},\bm{j}' $ with $ \bm{j}' \ne \bm{j} $.  $  H_{\mbox{\tiny on-site}} $ thus describes an array of independent QDs, while $ H_{\mbox{\tiny site-site}} $ contains the interactions between them.  In writing (\ref{onsitehamiltonian}), we have used the identity  $ {\cal T} (\bm{n},\bm{n}',\bm{j},\bm{j}') = \frac{\delta_{\bm{n} \bm{n}'}\lambda^2}{4\pi^2} \int_{{\bm{n}}\mbox{\tiny th B.Z.}} d^2 k E_{\bm{k} \bm{n}} e^{i \bm{k} \cdot (\bm{x}_{\bm{j}} - \bm{x}_{\bm{j}'})} $, where $ E_{\bm{k} \bm{n}} $ is the energy dispersion of the $ \bm{n}$th non-interacting band, and defined $ \epsilon_{\bm{n}} \equiv  {\cal T} (\bm{n},\bm{n},\bm{j},\bm{j}) $ and $ {\cal \tilde{U}}_{\bm{n}_1,\bm{n}_2,\bm{n}_3,\bm{n}_4} =  {\cal U} (\bm{j},\bm{j},\bm{j},\bm{j},\bm{n}_1,\bm{n}_2,\bm{n}_3,\bm{n}_4) $.  
$ H_{\mbox{\tiny on-site}} $  has precisely the same form as the Hamiltonian of an array of isolated QDs \cite{kouwenhoven01}. The only difference here is that the single particle basis used here
is a Wannier basis, instead of the eigenstates of the confining potential. For this reason, we henceforth call the many-body system of electrons interacting via Hamiltonian (\ref{onsitehamiltonian}) on a particular site a ``Wannier quantum dot'' (WQD).  In the limit of very large barriers between the dots, $ \epsilon_{\bm{n}}$ coincides exactly with the energy levels of an
isolated QD. 

\begin{figure}[t]
\scalebox{0.5}{\includegraphics{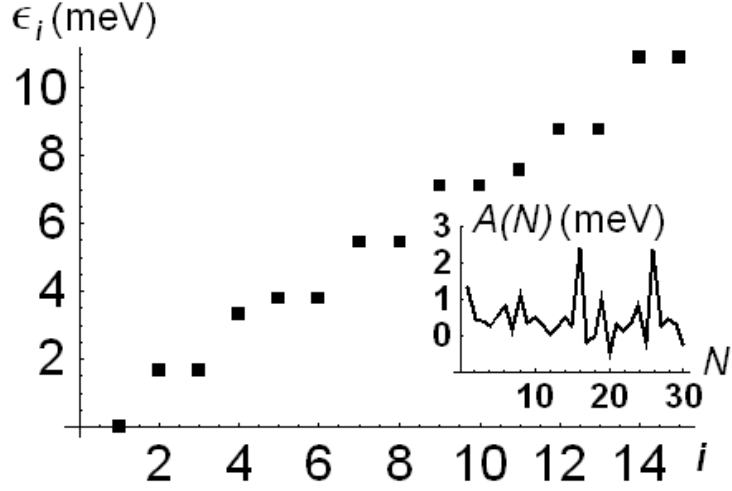}}
%\scalebox{0.4}{\includegraphics[width=\textwidth, bb=0 0 477 305]{spectrum.png}}
\caption{\label{fig:addition} The single particle energy spectrum $ \epsilon_i $ of a 2D Wannier quantum dot (WQD) versus
the energy level $ i $. The $i$th level can be occupied by two electrons due to spin. The inset shows the addition 
energy $ A(N) $ of a WQD.  We use parameters $ \lambda = 0.1 $ $\mu$m, $ V_0 = 1.1 $ meV, and $ d = 10 $ nm 
for the calculation.}
\end{figure}

In order to analyze the Hamiltonian (\ref{onsitehamiltonian}), we have performed an exact diagonalization study for electron
numbers up to $ N = 30 $.  In the inset of Fig. \ref{fig:addition} we show the electron addition spectrum calculated according to 
$ A(N)  = E_{N-1} + E_{N+1} - 2 E_N $ \cite{kouwenhoven01}, where $ E_N $ is the ground state energy of the $ N $-particle Hamiltonian.  We see a peak 
structure that reflects the shell structure of the WQD, with peaks in $ A(N) $ occuring at the magic numbers of the WQD, i.e. electron numbers corresponding to completely filled shells.  The 
spectrum of a single electron WQD is shown in the main plot of Fig. \ref{fig:addition} to show the shell structure.  Unlike
normal QDs, the shells are typically only doubly degenerate at most. The two-fold degeneracy originates from the $x$-$y$ symmetry of the potential. 
For electron numbers in the WQD corresponding to the doubly degenerate shells, we have verified that Hund's rule holds according to our numerical calculations, with ground states occuring for a total $z$-component of spin $ S_z=\pm 1$, rather than $ S_z=0$ at half-filled shell fillings.

Using the diagonalized states of (\ref{onsitehamiltonian}), we may write down an effective Hubbard Hamiltonian. 
The approximation we make is that for a given electron density, only one shell of a WQD is relevant to 
describe the low energy physics.  For fillings corresponding to a non-degenerate shell,
we may make the state associations  $  |0 \rangle_{\bm{j}} \equiv | N_b,0, \bm{j} \rangle$, $|\uparrow \rangle_{\bm{j}}  \equiv | N_b+1,1/2, \bm{j} \rangle$,
$| \downarrow \rangle_{\bm{j}} \equiv | N_b+1,-1/2, \bm{j} \rangle$, $| \uparrow \downarrow  \rangle_{\bm{j}}  \equiv | N_b+2,0,\bm{j} \rangle  $
for each site $ \bm{j} $, where the eigenstates of (\ref{onsitehamiltonian}) on a single site $ \bm{j} $ are denoted $ |N, S^z ,\bm{j} \rangle $,  and the total number of electrons in the lower energy ``core'' shells 
is called the base electron number $ N_b $. For doubly degenerate shells, a maximum of four electrons may occupy the shell, 
which gives rise to a two-band Hubbard model.  In this paper, for simplicity we consider the non-degenerate case, 
although the generalization to the two-band case may be performed straightforwardly.  

\begin{figure}[t]
\scalebox{1}{\includegraphics{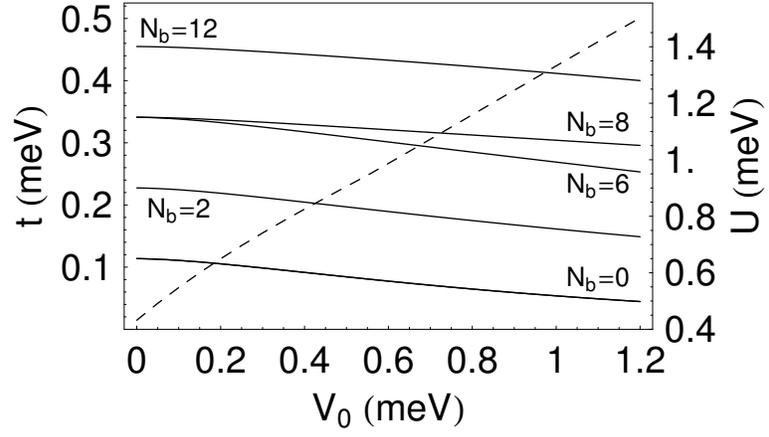}}
%\scalebox{0.4}{\includegraphics[width=\textwidth, bb=0 0 500 327]{figures/N1Uandt.png}}
\caption{\label{fig:effective_hopping}
The on-site Coulomb energy $ U $ (dashed line, right axis) for $ N_b = 0 $ and hopping $ t $ (solid lines, left axis) versus the periodic potential amplitude.  The base electron number for each $ t $ are labeled. Parameters  $ \lambda = 0.1 $ $ \mu $m and $ d = 30 $ nm are used for the calculation. }
\end{figure}

Let us now introduce a particle number operator $ n_{\bm{j} \sigma} $ acting on the states $ { \cal S}_{\bm{j}} =  \{ | 0 \rangle_{\bm{j}}, | \uparrow \rangle_{\bm{j}}, | \downarrow \rangle_{\bm{j}}, | \uparrow \downarrow \rangle_{\bm{j}} \} $ with the properties 
$ n_{\bm{j} \sigma} | 0 \rangle_{\bm{j}'}  = 0 $, $ n_{\bm{j} \sigma} | \sigma' \rangle_{\bm{j}'}  = \delta_{\bm{j} \bm{j}'} \delta_{\sigma \sigma'}| \sigma' \rangle $, 
$ n_{\bm{j} \sigma} | \uparrow \downarrow  \rangle_{\bm{j}'} = \delta_{\bm{j} \bm{j}'}| \uparrow \downarrow  \rangle $. 
In the space of $ {\cal S}_{\bm{j}} $, we may write down an effective Hamiltonian
\begin{equation}
\label{effectiveonebandham}
H^{\mbox{\tiny eff}}_{\mbox{\tiny on-site}}  =\sum_{\bm{j}} ( E_0 + 
\sum_{\sigma=\pm 1/2} \mu_\sigma n_{\bm{j} \sigma}
+ U n_{\bm{j} \downarrow} n_{\bm{j} \uparrow} )
\end{equation}
where $ E_0 = E_{N_b,0}$, $ \mu_\sigma = E_{N_b+1,\sigma} - E_{N_b,0} $, $ U = E_{N_b+2,0} - \sum_{\sigma=\pm 1/2} E_{N_b+1,\sigma} + E_{N_b,0} $.  The $ \mu_\sigma $ is an 
effective chemical 
potential term, while $ U $ is an effective Coulomb on-site repulsion energy. 
Figure \ref{fig:effective_hopping} shows the on-site repulsion $ U $ as a function of the 
periodic potential amplitude for a base electron filling $ N_b = 0 $. 
We see that the periodic potential amplitude increases the effective 
on-site interaction, in agreement with previous calculations \cite{byrnes07}. 

The hopping terms between sites are calculated according to the transition that the inter-site Hamiltonian $ H_{\mbox{\tiny site-site}} $ induces between the WQD eigenstates by making
a transformation of the hopping terms in $ H_{\mbox{\tiny site-site}} $ into the truncated basis states  $ { \cal S}_{\bm{j}}$. To a  good approximation we find that the hopping can be approximated by 
the hopping integral of the band corresponding to the outermost shell of the non-interacting WQD. 
The nearest neighbor hopping amplitudes for various base fillings $N_b$ are plotted in Fig. \ref{fig:effective_hopping}. We see that 
as the periodic potential is increased, the hopping is suppressed, due to the increased strength of the barriers between the QDs. 
As the number of electrons is increased, the tunneling is enhanced due to the electrons occupying higher energy bands.

Table \ref{tab:Utmin} gives our calculated values of various Hubbard parameters for two potential amplitudes 
$ V_0 = $ 0.56 and 5.4 meV and various base electron numbers $ N_b $. 
 For a given potential $V_0 $, the nearest neighbor and next-nearest neighbor hopping amplitudes $ t $ and $ t' $ increase with the 
base filling $ N_b $. 
Naively one may conclude based on this that using high base electron numbers is 
always advantageous due to the larger energy scale of the effective Hamiltonian. However, one must also consider whether inter-band transitions are suppressed sufficiently in order that the single band (or two-band) approximation is valid. For low temperatures $ k_B T \ll \frac{\hbar^2}{2 m^* \lambda^2} $, an approximate criterion is when there is no energy overlap between a given energy band and all other bands for the non-interacting band spectrum. 
For $ V_0 = 0.56 $ meV, a simple calculation reveals that only the lowest band is separated in energy from the other bands, hence an effective Hubbard model will only be formed for $ N_b = 0 $.  For $ V_0 = 5.4 $ meV, we find that the four lowest energy bands are separated, hence we require $ N_b <12 $. 
Since both $ V_0 $ and the average electron density are freely choosable parameters, in practice we may always obtain an effective Hubbard model by appropriately choosing these values.

By choosing a band with a small $ U/t $ and increasing the periodic potential $ V_0 $, we expect a metal-Mott insulator transition to occur in the system \cite{byrnes07,imada98}.  In our proposed device, such a transition may be identified by measuring the zero-bias differential conductance across the source and drain contacts. The conductance as a function of the chemical potential has been theoretically investigated in previous studies \cite{stafford94}. In the metallic state, we expect the system to be conducting when the Fermi level lies within a band created by the 
effective Hubbard model.  In the Mott insulating limit, the spectrum should reduce to that of a large array of isolated QDs. The 
characteristic conductance spectra of the metallic and insulating states differ in each case, thus by varying the potential $ V_0 $
at a fixed Fermi level, one should be able to distinguish a transition between the two phases. Magnetoconductance measurements provide an independent check  of a QPT between the metallic and Mott insulating states. The
phase-breaking length will reduce to a length comparable to the QD size
in the case of a Mott-insulating QD array \cite{liang94}. 
Experimental data consistent with these expectations have already been observed to some extent \cite{kouwenhoven90,haug92}. However, a clear metal-Mott insulator transition has never been identified to our knowledge. 

In order that such QPT phenomena are not washed out due to the temperature effects, 
we require the Hubbard parameters to be larger than the thermal energy $ U, t \gg k_B T $ \cite{sachdev99}. Using dilution refrigerator techniques, temperatures of $ T = 10 $ mK are reachable, corresponding to a thermal energy of $ k_B T \approx 1 $ $\mu$eV.  Previous theoretical estimates have suggested that 
temperatures of $ T \approx 0.1 t $ are necessary to observe an antiferromagnetic (AF) phase, and $ T \approx 0.02 t $ for a $d$-wave superconducting (SC) phase at $U = 4t $ \cite{hofstetter02,maier05}. Assuming these numbers, we require a hopping of about $ t > 0.01 $ meV for observation of an AF phase and $ t > 0.04 $ meV for observation of a SC phase.  Comparison of these numbers with Table \ref{tab:Utmin}, we find that both these phases should be possible to observe within the Hubbard approximated bands. The AF nature of the insulating
phase can be determined from temperature-varying magnetic
susceptibility measurements \cite{ashcroft76}. Evidence of Cooper-pair formation may be obtained
from the magnetocapacitance oscillation period, by observation of the 
Cooper pair charge of $2e$ in the strongly coupled
QD array. 

In summary, we have proposed an experimentally viable quantum simulator for the one- and two-band Hubbard models
using a semiconductor QD array device. For a given average electron number in the QDs, the low-energy physics may be described by an effective one- or two-band Hubbard model. Our scheme may be easily generalized to different lattice geometries simply by adjusting the mesh gate design and voltage.  Examining the region $ U/t \gg 1 $ produces an effective $t$-$J$ or Heisenberg model, while spin models involving frustration may also be explored by fabricating a triangular lattice mesh gate.  Another possibility is to introduce controlled disorder into the system by randomly varying the mesh dimensions in the lattice, and thereby producing a Hubbard-Anderson model.

This work is supported by JST/SORST, NTT, the University of
Tokyo, and the Special Coordination Funds for Promoting Science
and Technology.
We would like to thank M. Beasley, I. Fisher, M. Jura, M. Topinka, J. Berengut, and P. Recher for helpful discussions.

%%%%%%%%%%%%%%%%%%%%%%%%%%%%%%%%

\begin{table}[ht]
\caption{\label{tab:Utmin}
(a)
The on-site Coulomb energy $ U $, the nearest-neighbor Coulomb energy $V$,  nearest-neighbor hopping $t$, and next nearest-neighbor hopping $t'$ are computed for various base filling numbers $N_b$ for a potential amplitude $V_0 = 0.56 $ meV. The nearest neighbor Coulomb interaction $ V $ is calculated using the standard single band approximation. Numbers in brackets denote errors on the last significant digit. The band index $\bm{n}$ corresponding to the filling level of the non-interacting Wannier quantum dot is given.  For base fillings corresponding to a two-band Hubbard model, we take the average on-site Coulomb interaction between the electrons species.
(b) As for (a) but for $V_0 = 5.4 $ meV.  }

\begin{tabular}{ccccccc}
\cline{2-7}
&$ N_b $  &  $ \bm{n} $ & $ U $ (meV)  & $ V $ (meV) & $ t $ (meV) & $ t' $ (meV) \\
\cline{2-7}
&0		&	(1,1)				& 0.95(3)		& 0.24(7)				&		0.080	&	0.011 \\
&2 	&	(1,2) (2,1)	& 0.8(3)		& 0.21(4)				&		0.192	&	0.028 \\ 
(a) &6 	&	(2,2)				& 1.5(10)		& 0.14(3)				&		0.305 &	0.044 \\
&8 	&	(1,3) (3,1)	& 1.3(10)		& 0.24(3)				&		0.323 &	0.021 \\
&12 	&	(2,3) (3,2)	& 1.3(5)		& 0.5(4)				&		0.435 &	0.037 \\
&16 	&	(1,4) (4,1)	& 1.0(5)		& 0.6(4)				&		0.438 &	0.020 \\
\cline{2-7} \\
\cline{2-7}
&$ N_b $  &  $ \bm{n} $ & $ U $ (meV)  & $ V $ (meV) & $ t $ (meV) & $ t' $ (meV) \\
\cline{2-7}
&0		&	(1,1)				& 4.3(1)		& 0.17(7)	 		  &		0.0016	&   $2.2 \times 10^{-6} $	\\
&2 	&	(1,2) (2,1)	& 3.2(2)		& 0.10(3)				&		0.020		& $ 4.8 \times 10^{-4} $\\ 
(b)&6 	&	(1,3) (3,1)	& 2(1)				& 0.30(6)				&		0.129 & 0.016 \\
&10 	&	(2,2)			  & 2(1)				& 0.05(3)				&		0.038 & $ 9.7 \times 10^{-4} $	\\
&12 	&	(1,4) (4,1)	& 4(2)				& 0.3(1)				&	  0.32 	& 0.035	 \\
&16 	&	(2,3) (3,2)	& 2(2)				& 0.10(3) 			&	  0.15 	& 0.017	 \\
\cline{2-7}
\end{tabular}
\end{table}

\end{document}